\documentclass{emulateapj}
\usepackage{amssymb,amsmath,graphicx}
\usepackage[T1]{fontenc}
\usepackage{epstopdf}

\newif\ifAMStwofonts
\AMStwofontstrue

\def\arcdeg{\hbox{$^\circ$}}
\def\slantfrac#1#2{\hbox{$\,^#1\!/_#2$}}
\def\kms{km~s$^{-1}$}
\def\degree{$^{\circ}$}

\def\ga{\mathrel{\hbox{\rlap{\hbox{\lower4pt\hbox{$\sim$}}}\hbox{$>$}}}}
\def\la{\mathrel{\hbox{\rlap{\hbox{\lower4pt\hbox{$\sim$}}}\hbox{$<$}}}}

\shorttitle{Turbulent Properties of MBM16}
\shortauthors{Pingel et al.}
\begin{document}

\title{Characterizing the Turbulent Properties of the Starless Molecular Cloud MBM16} 

\author{Pingel, N.M.\altaffilmark{1}, Snezana Stanimirovi{\'c}\altaffilmark{1},
J.E.G. Peek\altaffilmark{2,}\altaffilmark{3},Min-Young Lee\altaffilmark{1},Alex Lazarian\altaffilmark{1},Blakesley Burkhart\altaffilmark{1}, Ayesha Begum\altaffilmark{4}, Kevin A. Douglas\altaffilmark{5}, Carl Heiles\altaffilmark{6}, 
Steven J. Gibson\altaffilmark{7}, Jana Grcevich\altaffilmark{2}, Eric J. Korpela\altaffilmark{6}, 
Allen Lawrence\altaffilmark{1}, Claire Murray\altaffilmark{1}, Mary E. Putman\altaffilmark{2}, Destry Saul\altaffilmark{2}}

\altaffiltext{1}{Department of Astronomy, University of Wisconsin-Madison, 
475 North Charter Street, Madison, WI 53706; pingel@astro.wisc.edu and sstanimi@astro.wisc.edu} 
\altaffiltext{2}{Department of Astronomy, Columbia University, New York, NY 10027}
\altaffiltext{3}{Hubble Fellow}
\altaffiltext{4}{Indian Institute of Science Education and Research, ITI Campus (Gas Rahat) Building, 
Govindpura, Bhopal- 23, India}
\altaffiltext{5}{University of Calgary/Dominion Radio Astrophysical Observatory,
P.O. Box 248, Penticton, BC V2A6J9, Canada}
\altaffiltext{6}{Astronomy Department and Space Science Laboratory, University of California, Berkeley, CA, 94703}
\altaffiltext{7}{Department of Physics and Astronomy, Western Kentucky University, Bowling Green, KY 4201}

\begin{abstract}
We investigate turbulent properties of the non-star-forming, 
translucent molecular cloud, MBM16 by applying the statistical technique of 
a two-dimensional spatial power spectrum (SPS) on the neutral hydrogen (H{\small I}) observations
obtained by the Galactic Arecibo L-Band Feed Array H{\small I} (GALFA-H{\small I}) survey. 
The SPS, calculated over the range of spatial scales from 0.1 to 17 pc, is well represented
with a single power-law function, with a slope ranging from -3.3 to -3.7 and being consistent over the velocity range of MBM16
for a fixed velocity channel thickness. However, the slope varies significantly with the
velocity slice thickness, suggesting that both velocity and density contribute to
H{\small I} intensity fluctuations. By using this variation we estimate the slope of 3D density fluctuations
in MBM16 to be $-3.7\pm0.2$. This is significantly steeper than what has been found for H{\small I} in the Milky Way plane,
the Small Magellanic Cloud, or the Magellanic Bridge, suggesting that 
interstellar turbulence in MBM16 is driven on scales $>17$ pc and 
that the lack of stellar feedback could be responsible for
the steep power spectrum.

\end{abstract}


\section{Introduction}

Interstellar turbulence is expected to play a key role 
in the formation and evolution of structure in the interstellar medium (ISM) \citep{mco07}. 
Numerous observational studies have shown that the ISM is turbulent on a vast 
range of scales. For instance, in the Galaxy and Magellanic Clouds, 
the hierarchy of spatial scales from smaller than a few parsecs 
to a few kiloparsecs  has been observed and often 
 attributed to interstellar turbulence \citep{cad83,gree93,sta99,dic00}. 
While in our current understanding many astrophysical processes 
require and depend on interstellar turbulence, many basic questions 
regarding the origin and properties of interstellar turbulence are still open. 
For example, what are the primary energy sources that induce 
turbulence over a broad range of scales? 
On what scales and through which modes is turbulent energy dissipated throughout 
the ISM? What properties of the interstellar gas determine the level and type of
turbulence within the ISM (e.g. magnetic fields, galaxy mass via tidal effects, gas density and porosity)? 

Statistical techniques for data analysis are very good at
modeling the general properties of complex systems.
In particular, the spatial power spectrum (SPS) has been frequently used
for decomposing observed images and measuring power/intensity associated with individual scales. 
SPS was applied to neutral hydrogen (H{\small I}) images for regions within our Galaxy \citep{cad83,gree93,dic01}, 
for the Small Magellanic Cloud, SMC, \citep{sta99,sal01}, the Large Magellanic Cloud, LMC, \citep{elm01} 
and the Magellanic Bridge \citep{mul04}. The results of these studies have shown 
the existence of a hierarchy of structures in the diffuse ISM, with the spectral slope
varying slightly with interstellar environments.
For example, H{\small I} integrated intensity images of several regions in the Milky Way, which are largely
confined to the disk and likely affected by high optical depth,
show a power-law slope of the SPS of $\sim-3$ (Lazarian \& Pogosyan 2004). 
A slightly steeper slope of $-3.4$ was found for the H{\small I} column density in the SMC, and $-3.7$ for the LMC.
The Magellanic Bridge slope (for the integrated H{\small I} intensity) varies across different 
regions between $-2.9$ and $-3.5$ (Muller et al. 2004). 
Integrated intensity maps of several gas and dust tracers have a power-law slope 
of $-3$ to $-4$ in the case of M33 (Combes et al. 2012).
Galaxies like the LMC and M33 show a break in their power spectra, while no departure from a single power law was
observed in the Galaxy, SMC and the Magellanic Bridge even at the largest probed spatial scales.

While main contributors to the interstellar turbulence and their 
relative importance are still
not well understood, it is expected that shear from galactic rotation
and large-scale gravitational/accretion flows could be responsible for
large-scale energy injection in galaxies, while stellar sources 
(stellar outflows, winds and jets, 
supernovae explosions and superbubbles) inject energy on smaller 
scales (McCray \& Snow 1979). 
As suggested by many numerical simulations (e.g. Vazquez-Semadeni et al. 2010), 
stellar feedback is expected to have a significant effect on cloud
evolution resulting in an increased velocity dispersion and a low star formation efficiency.
In addition, numerical simulations are starting to use statistical properties of the ISM (e.g. Pilkington et al. 2011)
to compare simulated galaxies with observations and constrain detailed feedback processes.
Clearly, observational constraints on the  stellar feedback and its influence on 
the statistical properties of the ISM are 
highly desirable, yet still rare.

In this paper we investigate the importance of stellar sources for 
turbulent energy injection by  studying the turbulent 
properties of the neutral hydrogen (H{\small I}) in the high-latitude ($\left|b\right|$ $>$ 30\degree)
molecular cloud MBM16 that is known to be absent of stars. We then compare our results 
with turbulent properties of the active stellar environments such as the SMC.
Our study is largely enabled by the high resolution H{\small I} observations 
provided by the GALFA-H{\small I} survey (Stanimirovi{\'c} et al. 2006, Peek
et al. 2011) which provide a wide range of spatial scales
essential for applying statistical techniques. 
	
MBM16 was first discovered by Magnani, Blitz \& Mundy (1985) through CO observations at 2.6 mm.
It is a large non-starforming (Magnani, LaRosa \& Shore 1993), translucent 
(mean $A_V=0.9$ mag, Hobbs et al. 1988) cloud 
at a distance of $\sim80$ pc \citep{hob88}. 
The cloud distance was bracketed (60 to 95 pc) 
using interstellar Na I absorption lines in the direction of stars located
behind the cloud and at varying distances. MBM16 has a mass of 320 $\pm$ 190 $M_{\odot}$ 
and is not gravitationally bound \citep{tmh98}. The lack of an external pressure 
confinement to counteract gravity suggests that  the cloud's lifetime is on the order of 
one Myr, which agrees with the lack of evidence for a stellar environment.
The only claim of potentially stellar component associated with MBM16 is from ROSAT X-ray observations
by Li, Hu \& Chen (2000) who identified four potential T Tauri star candidates and concluded that future work is needed
to confirm this association.
In conclusion, numerous studies suggest that MBM16 has a relatively 
quiescent interstellar environment 
not polluted with current star formation.

Besides being mapped in CO (Magnani, Blitz \& Mundy 1985, LaRosa, Shore \& Magnani 1999) and H{\small I}
(Gir, Blitz \& Magnani 1994), several clumps of formaldehyde (H$_2$CO) were detected in absorption
by Magnani, LaRosa \& Shore (1993) suggesting a density of a few times $10^3$ cm$^{-3}$ at most.
Regarding its large-scale environment, MBM16 is located about 10 degrees away from the 
Taurus molecular cloud. While several studies suggested that MBM16 could be associated with 
the Taurus-Auriga-Perseus complex of molecular clouds (based
on the similar LSR velocity and morphology in IR images), 
this association is questionable considering that
MBM16 is at a distance of about half that of Taurus \citep{hob88}.
In general, several previous studies suggested that high-latitude  translucent molecular clouds 
were formed in regions compressed with shocks and therefore are expected at intersections of larger
gaseous (shells, sheets, flows etc) structures.

Turbulent properties of molecular gas in MBM16 were a subject of several previous  studies.
Magnani, LaRosa \& Shore (1993) suggested that the observed properties of H$_2$CO clumps 
could be explained as a result of a large-scale shear instability. 
This idea was further developed in 
LaRosa, Shore \& Magnani (1999) who calculated the autocorrelation function of the
$^{12}$CO  velocity field.
They were able to determine a velocity correlation scale of 0.4 pc 
and attributed this to the formation of coherent structure due to an external shear
 flow.  They pointed to this type of flow as a possible source for both 
magnetohydrodynamic (MHD) turbulence on small scales in both this and 
other molecular clouds. They also concluded that such flows could continuously supply kinetic energy
 to molecular clouds over long time scales.

More recently, Chepurnov et al. (2010) studied a high-latitude field spatially adjacent to MBM16 (RA 02$^{h}$ 00$^{m}$ to 02$^{h}$ 24$^{m}$,
Dec 7$^{\circ}$ to 13$^{\circ}$ in J2000) also using GALFA-H{\small I} observations. The H{\small I} emission in the studied region was predominantly at negative velocities
($-20$ to 0 \kms). 
They applied a new statistical method, the velocity coordinate spectrum, where a one-dimensional power spectrum
is derived along the velocity axis. A steep power-law slope of the velocity fluctuation
spectrum of $-3.9\pm0.1$ was derived, and a more shallow 
slope of $-3.0 \pm 0.1$ for the power spectrum of density fluctuations.
Chepurnov et al. (2010) interpreted these results as being caused by a shock-dominated turbulence.

In this paper, we apply the SPS analysis on the H{\small I} observations of MBM16 to complement previous 
statistical investigations, and provide a basis for future studies of high-latitude clouds from the GALFA-H{\small I} survey. 
The paper is structured as follows: $\S$ 2 summarizes the GALFA-H{\small I} observations and data reduction.
In $\S$ 3 we summarize the basic cloud properties.
In $\S$ 4, we derive the spatial power spectrum 
and investigate how its slope varies with velocity and velocity channel thickness, and 
we discuss our results in  $\S$ 5.

\section{Data}

The data used in this study come from the first data release of the GALFA-H{\small I} survey (DR1). 
GALFA-H{\small I} is a high resolution ($4'$), large area (13,000 deg$^{2}$), and high spectral 
resolution (0.18 km s$^{-1}$) survey conducted with the 7-beam array of receivers ALFA at the 
Arecibo\footnote{The Arecibo Observatory is operated by SRI International under a cooperative 
agreement with the National Science Foundation (AST-1100968), and in alliance with Ana G. 
Mendez-Universidad Metropolitana, and the Universities Space Research Association.} radio 
telescope.  Observations were obtained using  the special-purpose spectrometer GALSPECT for Galactic 
science. The region used for this study comes from a combination of 
``basket-weave'' observations (PI: Peek), where the telescope 
is fixed at the meridian and is driven up and down in zenith angle at chosen declination (Dec) 
range to minimize variation of instrumental effects, and commensal ``drift-scan'' observations 
(PIs: Putman \& Stanimirovi{\'c}, Heiles \& Stanimirovi{\'c}), where
scans at a constant declination are observed while the sky is drifting. 

Data reduction of H{\small I} observations is 
implemented using the GALFA-H{\small I} Standard Reduction (GSR) pipeline. 
DR1 covers 7520 square degrees of the sky and has been 
produced using 3046 hours of integration time from 12 different projects. 
DR1 data cubes are publicly available online and a second data release is in preparation. 
While full details of the 
GSR can be found in Peek et al. (2011), we emphasize that DR1 has
been calibrated for the first sidelobe contribution and further stray radiation correction
(due to more distant sidelobes) is planned for future data releases.
The estimated contribution of the stray radiation is $<1$ K for the region
of interest in this study (based on the comparison with well calibrated surveys). This 
has negligible effect on the results of our SPS analysis in Section 4 as the HI emission 
under investigation is very bright (e.g. Figure 5).

Our region of interest  extends from RA 2$^h$ 50$^m$ to 3$^h$ 45$^m$, Dec 4\degree to 16\degree (J2000).
The rms noise measured in the extracted data cube over emission-free channels is $\sigma$ = 0.43 K 
at the original velocity resolution of 0.18  km s$^{-1}$. Please note that in the SPS analysis 
(Section 4) where we average several velocity channels we re-calculate the rms noise before 
estimating the uncertainty for the HI column density.

\section{Observational Results}

\begin{figure*}
\epsscale{1.6}
\centering{\includegraphics[scale=.7]{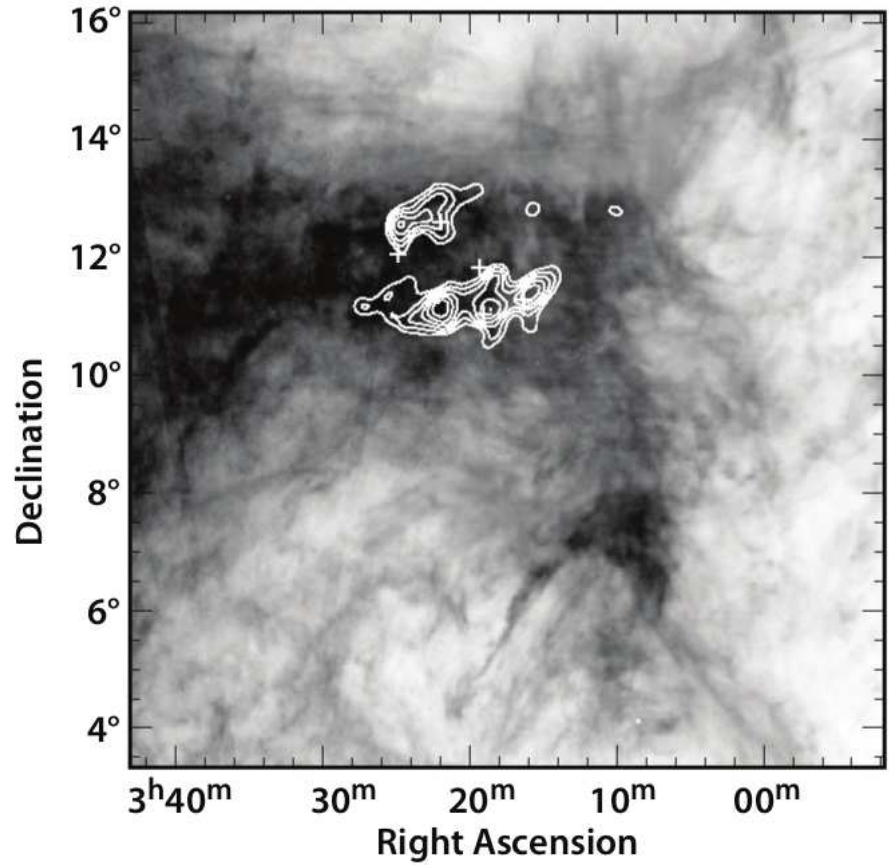}
\caption{An HI emission image of MBM16 at a velocity of 8 \kms (LSR).
The CO integrated intensity from  LaRosa, Shore, \& Magnani (1999) is shown with white contours 
from 40\% to 90\% with steps of 10\% of total intensity. Crosses are CO detections from Magnani et al. (2000).}
\label{fig:fig1}}
\end{figure*} 

\begin{figure*}
\epsscale{1.}
\centering{\includegraphics[scale=0.75]{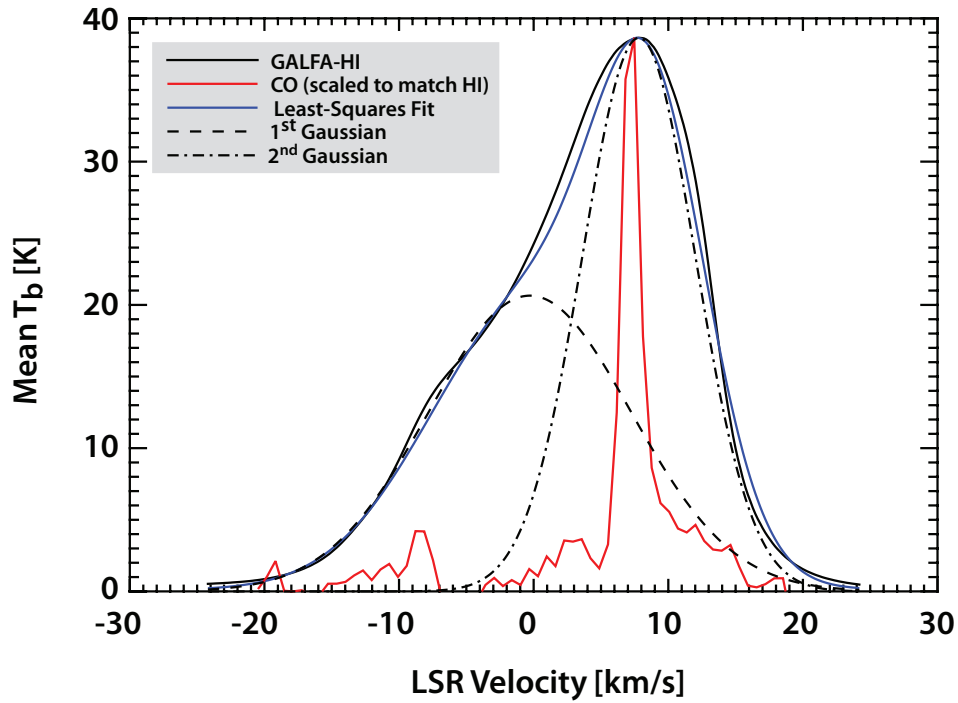}
\caption{The mean HI profile of MBM16 (black) and 
the scaled CO spectrum (red) highlighting the velocity range of the data set used. 
Two fitted Gaussian functions are shown as dashed lines and the combined fit is shown in blue.}
\label{fig:fig2}}
\end{figure*}

The region under investigation in this study encompasses the CO emission
from LaRosa, Shore, \& Magnani (1999) but is significantly more extended.
This is shown in Figure 1 where in grey-scale we show the H{\small I} emission at the velocity of the
peak CO intensity, with CO integrated intensity overplotted in white.
The H{\small I} emission essentially forms an arc-like feature, with the CO map covering
the arc's deflection point.
However, there are indications that CO may be more extended than what is shown by LaRosa, Shore, \& Magnani (1999).
We have added here as crosses six CO detections from \cite{magnani00} from an
unbiased CO survey of the southern Galactic hemisphere with the 1.2m CfA telescope. 
While four crosses are coincident with white contours,
two detections are located just outside the traditionally considered CO boundary of MBM16.
All of the H{\small I} emission shown in Figure 1 is included in our analysis.

The average H{\small I} profile for the whole area shown in Figure 1 is shown in Figure~\ref{fig:fig2} (black),
together with the  average, scaled $^{12}$CO (J = 1-0) profile from
LaRosa, Shore, \& Magnani (1999) (red). This shows that H{\small I} is more extended than CO, 
similar to what has been seen for many other molecular clouds.
This averaged H{\small I} profile appears to have two Gaussian components based on 
least-squares fitting. The stronger Gaussian component, with 
the peak brightness temperature of 39 $\pm$ 2 K,
agrees well with the molecular component in terms of its velocity center
(8 \kms) and range (4 to 10 km s$^{-1}$, FWHM of 9.9 $\pm$ 0.3 km s$^{-1}$). 
The smaller Gaussian component peaks at a brightness temperature of 20 $\pm$ 1 K and has a 
FWHM of 16.4 $\pm$ 0.8 km s$^{-1}$.
LaRosa, Shore, \& Magnani (1999) also noticed 
a weaker molecular cloud between $-20$ and $-5$ km s$^{-1}$. While the distance of this second cloud is not known, 
it could be associated with the second (weaker) Gaussian H{\small I} component.

\begin{figure*}
\epsscale{0.6}
\centering{\includegraphics{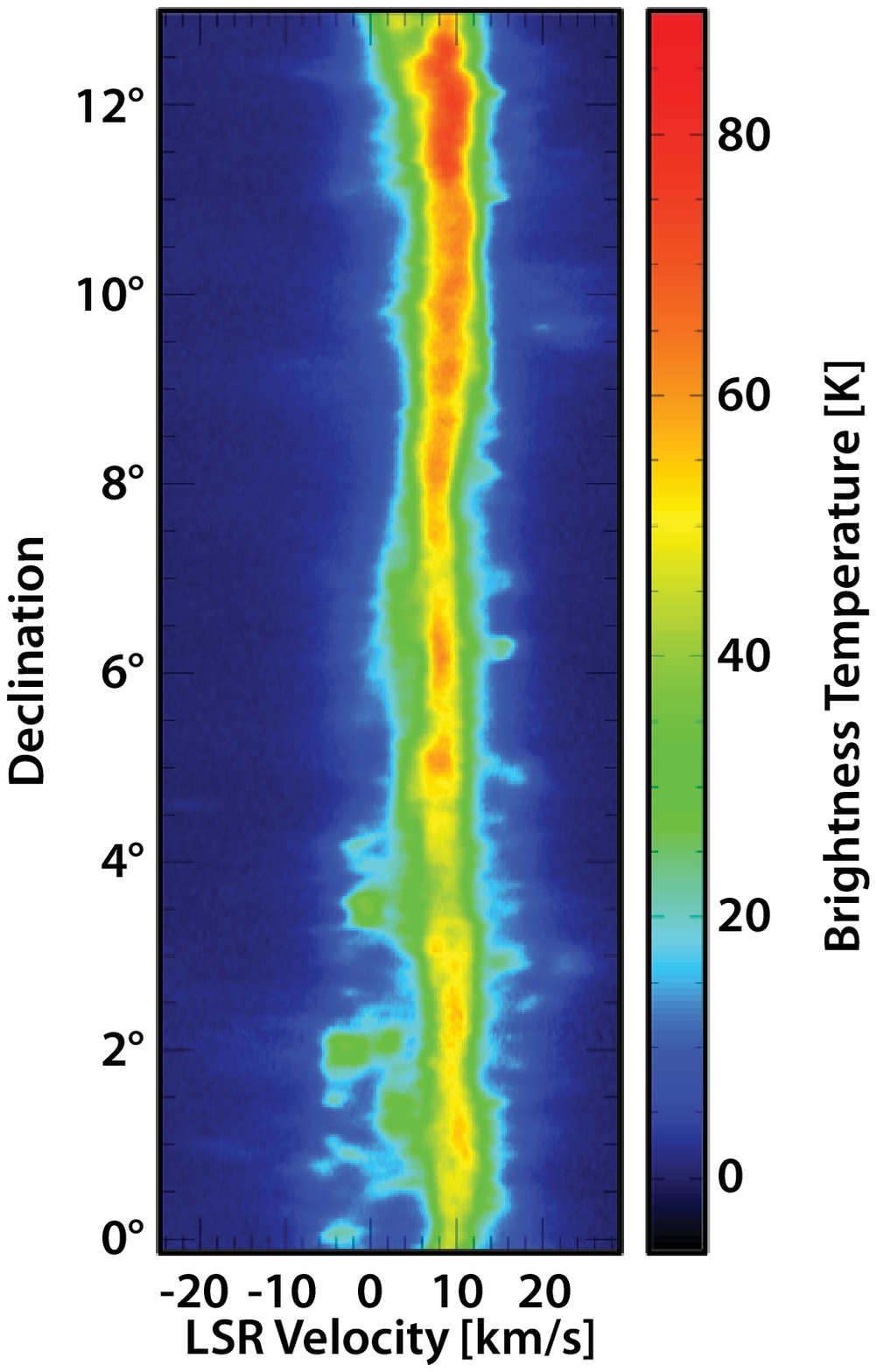}
\caption{The integrated intensity image along the RA axis of the region shown in Figure 1. HI between $-5$ and $-10$ km s$^{-1}$ may be associated with a separate molecular cloud \citep{lsm99} located at a different distance than MBM16.}}
\label{fig:fig3}
\end{figure*}

In Figure~\ref{fig:fig3} we show an image obtained after integrating the H{\small I} 
data cube along the RA axis calculated 
over the whole region of interest.
While a small amount of gas is present at negative LSR velocities, across most of the spatial extent
the main velocity component dominates and is well-centered at $\sim8$ \kms.
The negative-velocity component is not obvious in this image as it has a localized spatial extent. It 
is centered at about five degrees south of the center of the CO emission in a cloud-like 
feature about 2 degrees in diameter.
To ensure that all velocity channels used for SPS analysis in the next section are filled with H{\small I} emission
we restrict the velocity range for our study to $-2$ to 16 km s$^{-1}$.  This range 
corresponds to the full width at half maximum of the integrated velocity profile in Figure 2.
We note that repeating our SPS analysis on the full velocity range of $-20$ to 20 \kms does not
result in significant differences.

\begin{figure*}
\epsscale{0.6}
\centering{\includegraphics[scale=.8]{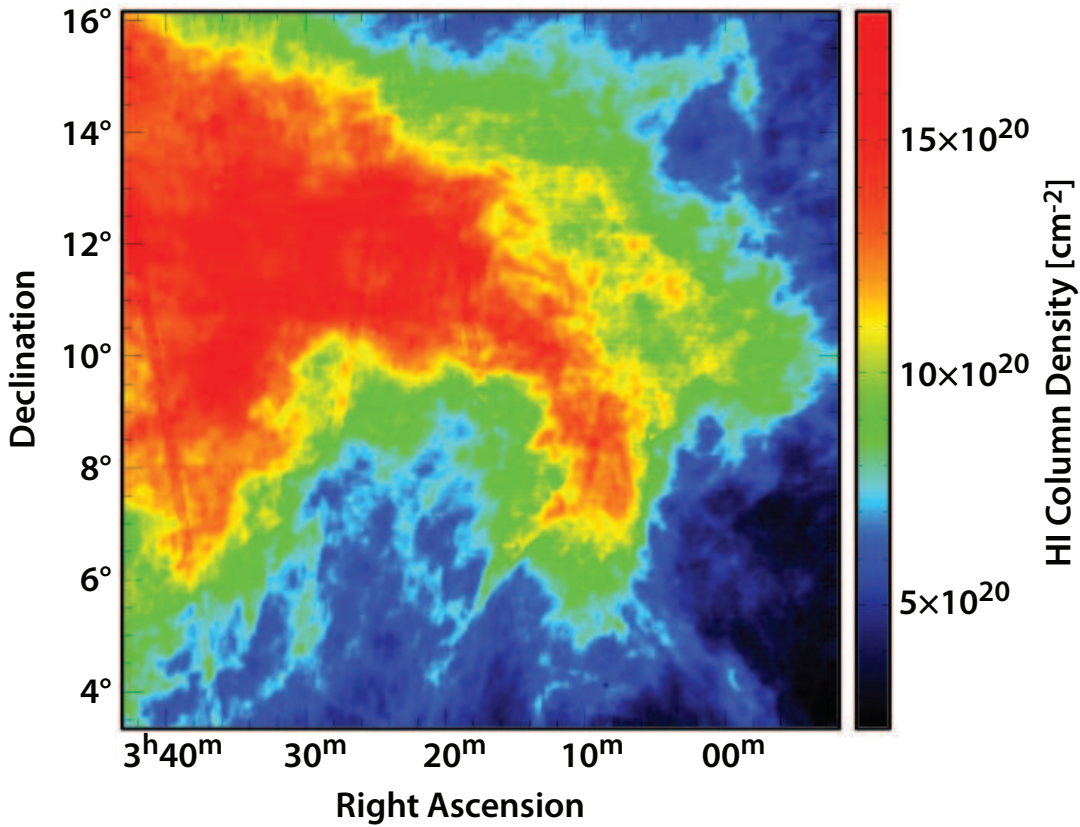}
\caption{The HI column density image of MBM16 in units of (cm$^{-2}$) 
integrated over $-2$ to 16 km s$^{-1}$. The physical scales range from 0.1 pc for the beam size to $\sim$17 pc for the size of the entire region. }
\label{fig:fig4}}
\end{figure*}

\begin{figure*}
\epsscale{1}
\centering{\includegraphics{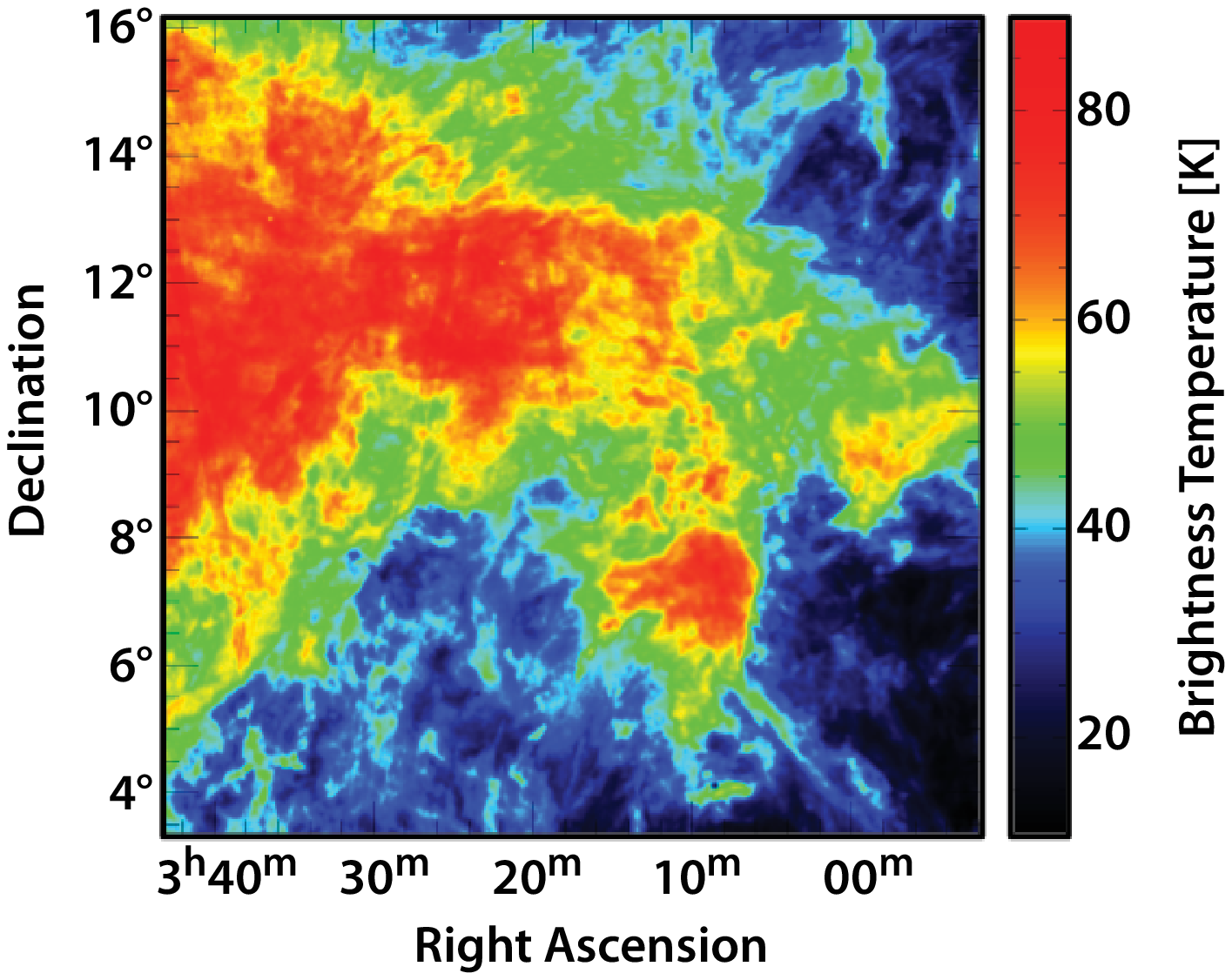}
\caption{The peak HI brightness temperature image calculated over the velocity range of $-2$ to 16 km s$^{-1}$. }
\label{fig:fig5}}
\end{figure*}

A column density image integrated over  $-2$ to 16 km s$^{-1}$ under the assumption of optically thin H{\small I} is shown in 
Figure~\ref{fig:fig4} with the minimum and maximum column density measurements 
being 4.7$\times$10$^{20}$ and 2.2$\times$10$^{21}$ cm$^{-2}$, respectively. 
Our derived mean H{\small I} column density of $9.3\pm 3.4\times 10^{20}$ cm$^{-2}$
agrees well with an average value of $9.3\times10^{20}$ cm$^{-2}$
measured by Gir, Blitz \& Magnani (1994) using the Green Bank 43 m telescope.
Figure~\ref{fig:fig5} shows a peak intensity image of H{\small I} associated with MBM16, 
and reveals the range of H{\small I} morphology of MBM16. Some regions, most notably the center, are 
smooth and continuous, while filamentary structure is present towards the edge of 
the image with sharp transitions between high and low brightness temperature values. 

\section{Spatial Power Spectrum }
\subsection{Derivation of the SPS}

\begin{figure*}
\epsscale{1}
\centering{\includegraphics{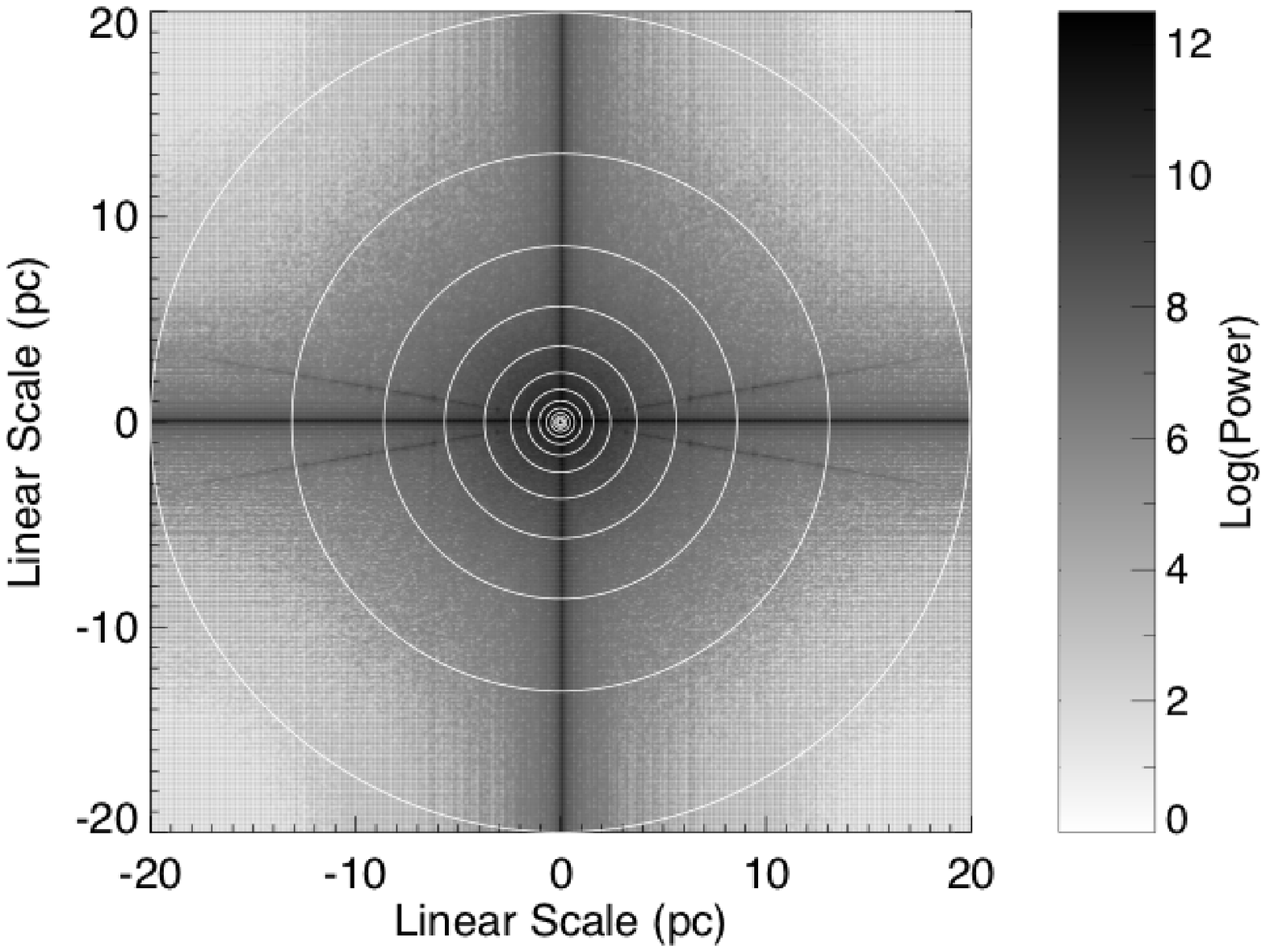}
\caption{An example modulus image, $<\Re^{2}+\Im^{2}>$, used to measure the spatial power spectrum.
Very bright pixels along the central axes are caused by sharp image edges, so called Gibbs phenomenon.
The bright lines inclined at an angle of about 15 degrees relative to the x-axis correspond to
the low-level imperfections in the gain calibration of our basket-weave scanning tracks (Section 4.1). 
}
\label{f:modulus}}
\end{figure*}

Our methodology for deriving the SPS is similar 
to previous studies \citep{cad83,gree93,sta99,sal01,elm01,mul04}. 
We begin with two-dimensional images of the H{\small I} intensity measurements ($I$). 
\citet{sal01} define the 2D spatial power spectrum as:
\begin{equation}
P(k)= \int \int\langle I\left(x\right)I\left({x}' \right )\rangle e^{i\bold L\cdot k}d\bold L, \ \bold L= \bold x-\bold {x}'
\end{equation}
with $\bold k$ being the spatial frequency, measured in units of wavelength and  
is $\propto$ 1/$l$ where $l$ is the length scale, and $\bold L$ being the 
distance between two points. 
This is essentially the Fourier Transform of the auto-correlation function of the H{\small I} intensity fluctuations.
We take the Fourier Transform of the intensity images and square the modulus of the transform --- 
$<\Re^{2}+\Im^{2}>$, where $\Re$ and $\Im$ are the real and imaginary components, respectively --- to obtain the power spectrum image, which we refer to as the modulus image. 
Thirteen annuli of equal width originating at the center of the modulus image are then 
overlaid at logarithmic intervals, as shown in Figure~\ref{f:modulus}. 
Each successive annulus corresponds to a certain 
linear scale in the intensity image. Larger annuli probe the amount of power at smaller spatial scales, while the smaller annuli probe the largest spatial scales within MBM16.  
To measure the average power per annulus we assume azimuthal symmetry and calculate the median value for each annulus.
The use of median is motivated by the fact that the modulus image has very 
bright pixels along the central axes caused by the Gibbs phenomenon, in which the Fourier series sum oscillates around discontinuities, and leads to $\sin(x)/x$ ringing resulting from the sharp image edges (Figure~\ref{f:modulus}).
While we have experimented with tapering of the original images so the intensity decreases 
smoothly to zero at the edges, the distribution of data values in each annulus is well behaved with 
the majority of data points following a Gaussian distribution and
ringing only introducing a high-end tail of the distribution.
Therefore, the average power per annulus is  well-represented with the median value.
This median power is then plotted as a function of spatial scale, as shown in Figures 7-8.

\begin{figure*}
\centering{\includegraphics[scale=.75]{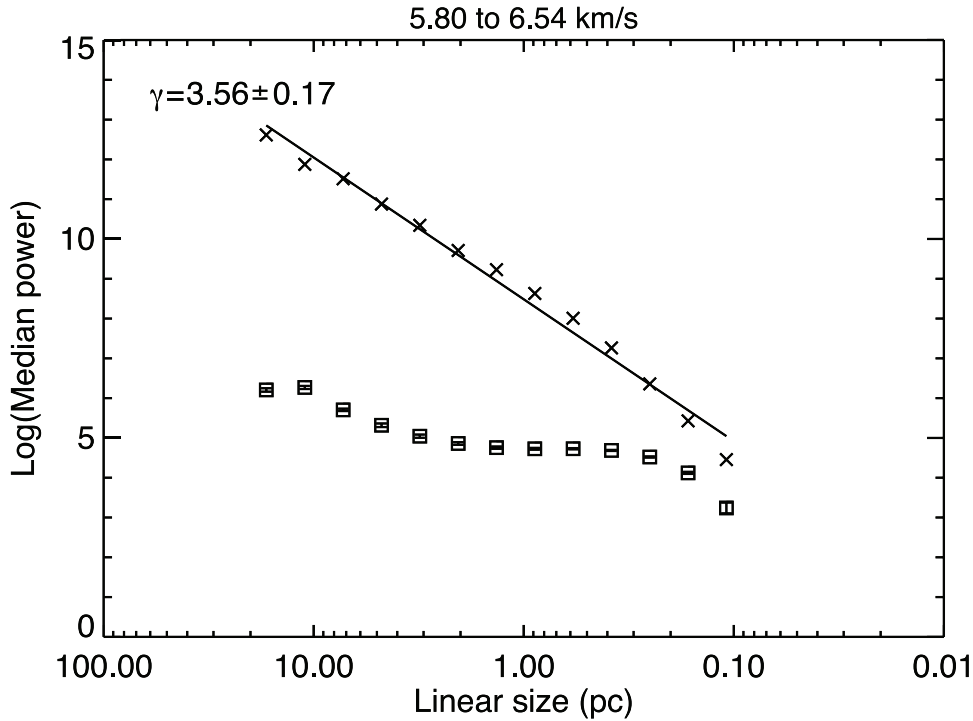}
\caption{An example power spectrum calculated for HI images over the velocity range of $5.8$ to $-6.5$ km s$^{-1}$. 
The power-law fit is overlaid.
Due to the small errors associated with calculating the median power, the error in the fit 
is determined by removing one data point at the time, re-fitting the power spectrum, and 
estimating the maximum deviation of the derived slopes.
The SPS shown with squares was derived using only emission-free channels in the data cube
over a velocity width of 0.72 km s$^{-1}$.}
\label{fig:fig6}}
\end{figure*}

To estimate the error in calculating the median power for each spatial scale 
we perform a Monte Carlo simulation. Based on our estimated uncertainty for image intensity (brightness 
temperature) per individual velocity channel,
we  generate 1000 H{\small I} intensity images by using 1000 intensity 
values for each image pixel, randomly drawn from a Gaussian distribution centered on the original 
pixel's value, and a standard deviation being equal 
$\sigma$ (where  $\sigma$ is the measured rms noise per
individual velocity channel). From here, for each annulus, we measure 1000 median values, and estimate
the final uncertainty as the standard deviation of these 1000 median values. 
These uncertainties range in value from four to two orders of magnitude 
below the calculated median power for a given spatial scale.
The uncertainties are shown in Figure~\ref{fig:fig6} 
but are generally too small to be noticed.
When integrating over several velocity channels in Section 4.2 
the standard deviation used in the Monte Carlo simulation is
calculated as: $\sigma \times \sqrt{N}  \times \Delta v$, where $N$ is the number of channels being averaged
and $\Delta$$v$ is the original velocity width of a single channel.

In addition to the above random uncertainties we also investigate the power spectrum of our emission-free
channels in the data cube. In the case that there are no systematic errors associated with temperature 
calibration and/or imaging, the noise power spectrum should be flat.  
According to \citet{gree93}, the observed power spectrum at a particular spatial  frequency $k$ 
 can be written as: 
\begin{equation}
P_{obs}\left(k\right)=P_{H\small I}\left(k\right)+P_{noise}\left(k\right)
\end{equation}
where P$_{H\small I}$ is the pure H{\small I} power spectrum and P$_{noise}$ is the 
noise-only contribution (assuming no absorption).
An example emission-free power spectrum is shown in  Figure~\ref{fig:fig6} using square symbols.
While this spectrum shows the existence of some residual power at larger-scales, likely caused
by imperfections in the gain calibration,
the noise power observed is 3-4  orders of magnitude smaller than the power measured in emission channels.
Therefore,  subtracting the noise power spectrum from 
the derived emission spectrum and re-fitting leads to a negligible change in the power-law slope of 0.1\%.
We therefore conclude that this residual noise mainly on scales 5-17 pc is not affecting our results.

\subsection{SPS Results}

\begin{figure*}
\centering{\includegraphics[width=3.2 in]{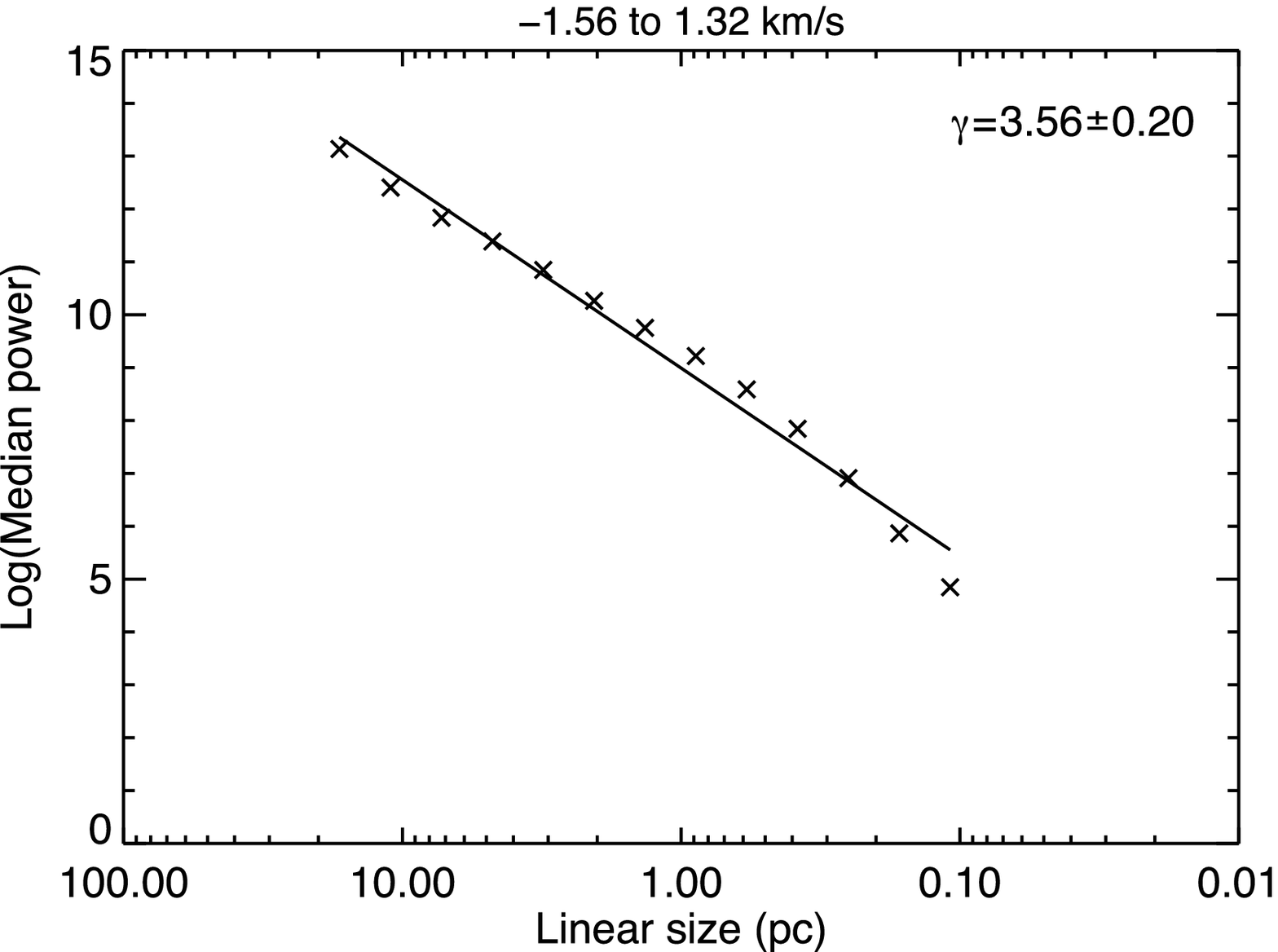}
\includegraphics[width=3.2 in]{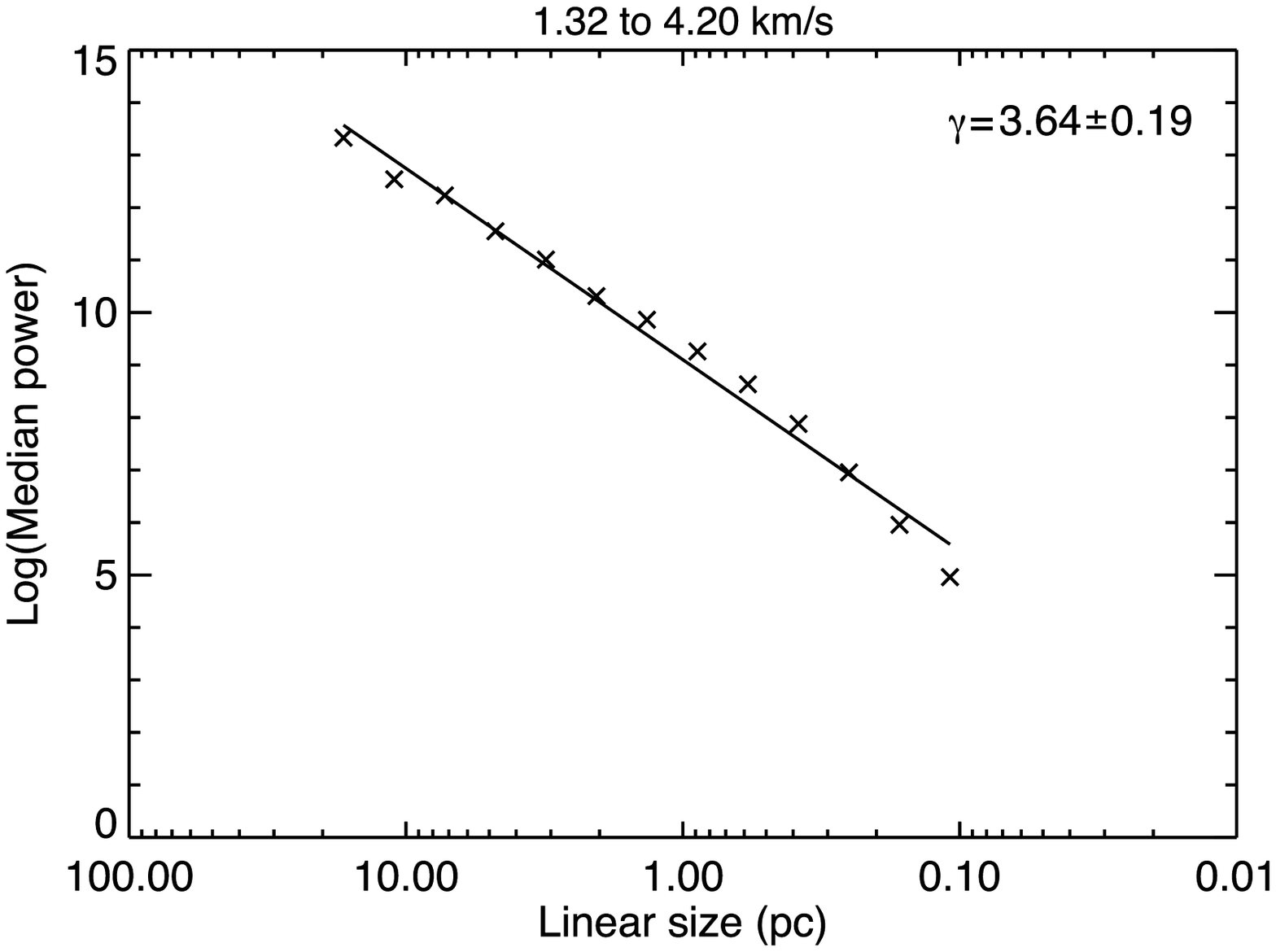}}
\centering{\includegraphics[width=3.2 in]{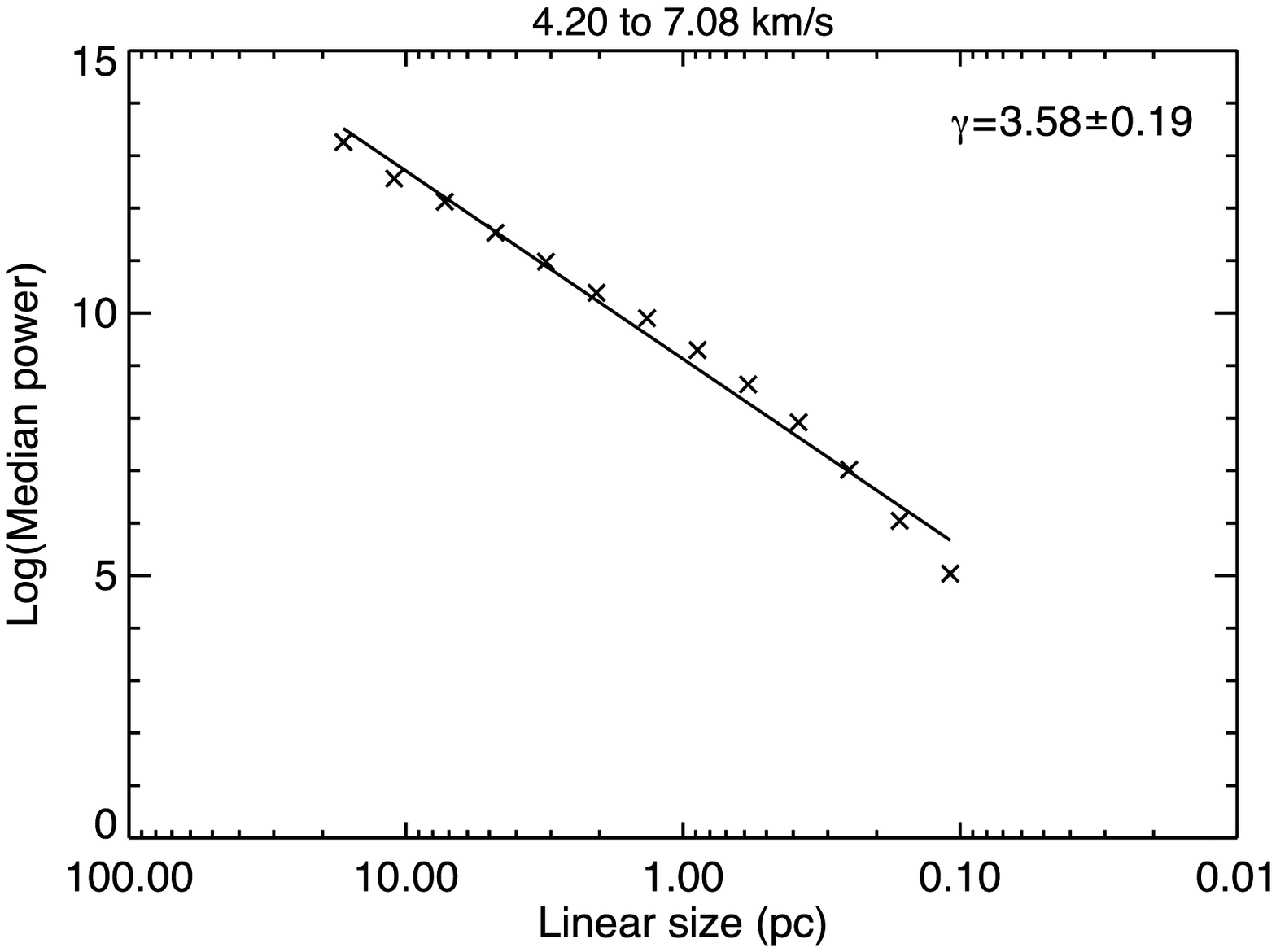}
\includegraphics[width=3.2 in]{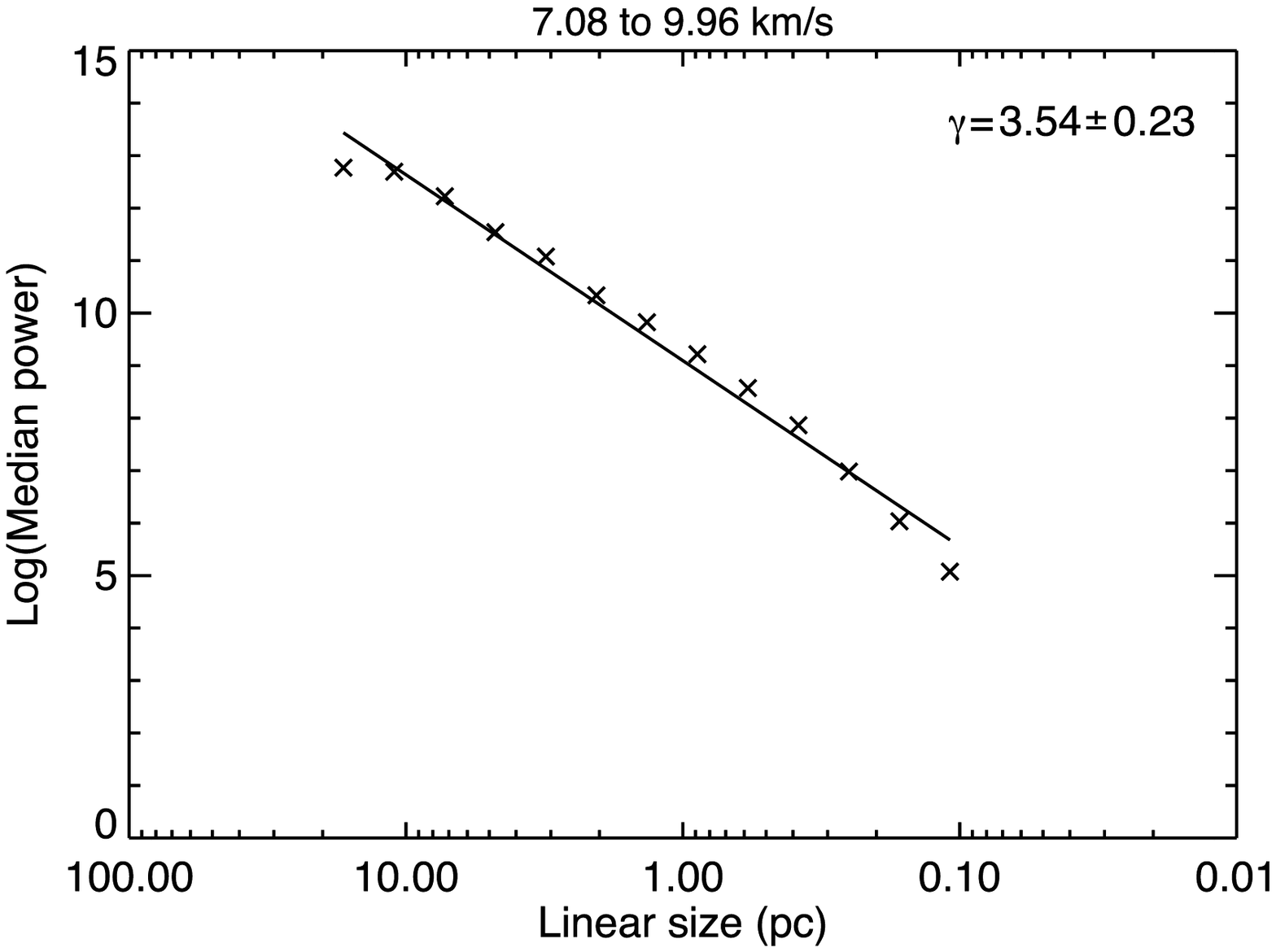}}
\centering{\includegraphics[width=3.2 in]{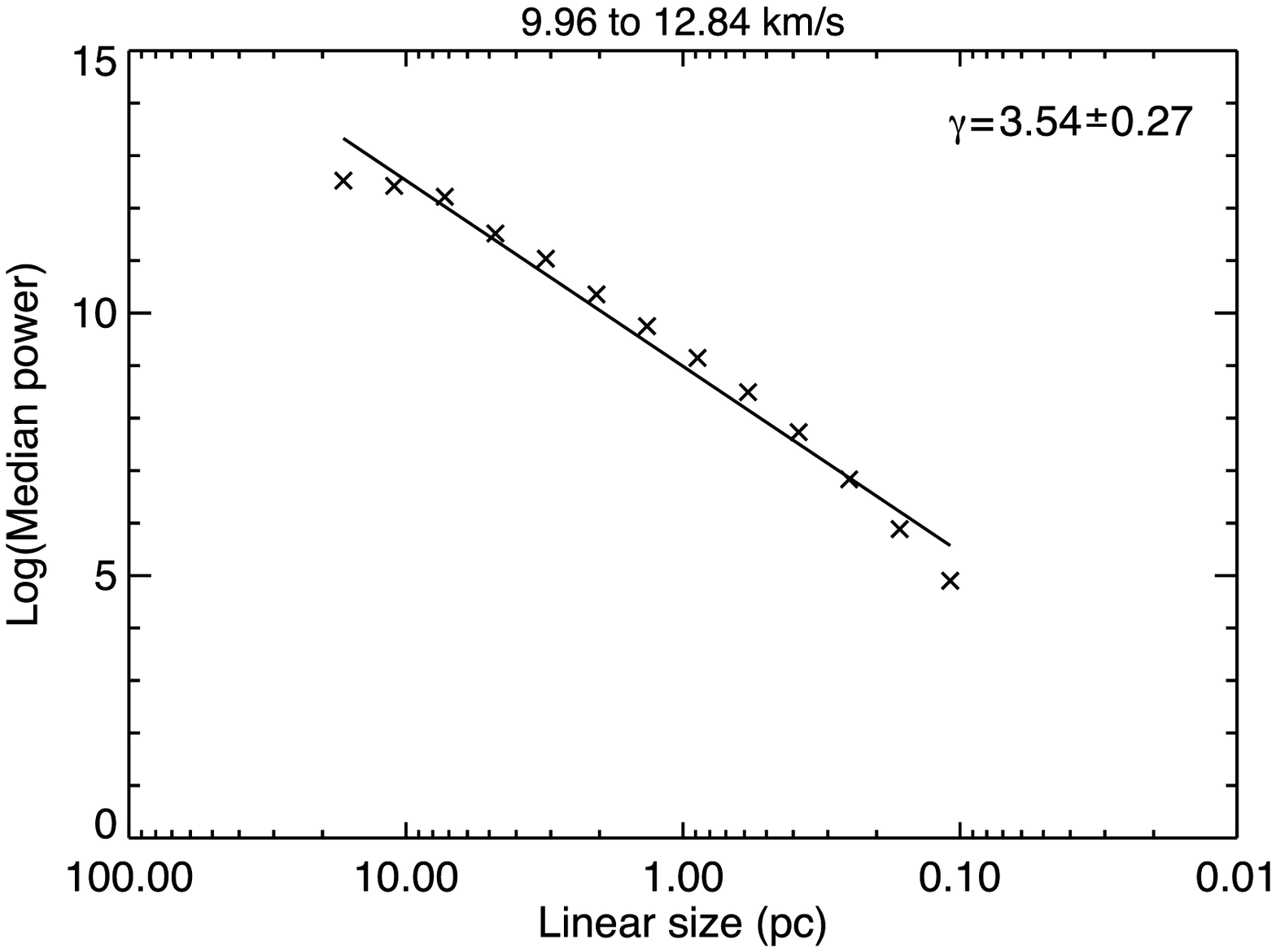}
\includegraphics[width=3.2 in]{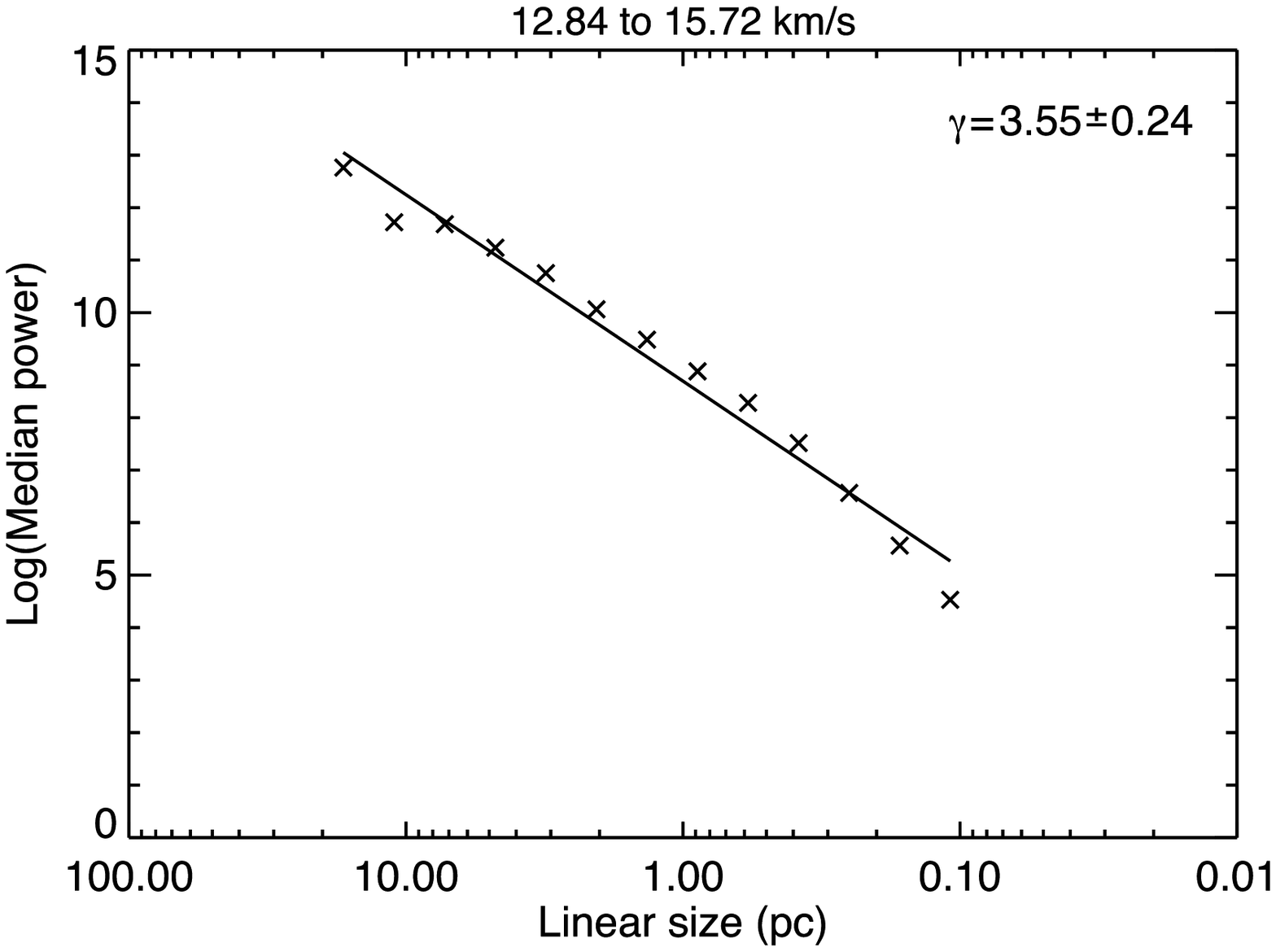}}
\caption{Example power spectra of velocity slices, each 
slice is the average of 16 velocity channels from the original HI data cube, with $\Delta$$v$=2.9 km s$^{-1}$.
The weighted mean of the six power-law slopes shown here is $-3.58\pm0.09$.}
\label{fig:fig7}
\end{figure*}

We begin by showing a few example power spectra calculated over the whole velocity range of interest 
 $-2$  to 16 km s$^{-1}$ for velocity channels averaged over 2.9 \kms (Figure~\ref{fig:fig7}), in order to obtain six equally wide velocity slices.
With the distance to MBM16 estimated to be approximately 80 pc \citep{hob88} 
and an angular resolution of the GALFA-H{\small I} survey 
equal to 4', the spatial frequency can directly relate to physical size. 
This is plotted on the x-axis and our observations cover a range of spatial scales from 0.1 pc to 17 pc.
The power spectra can clearly be fitted with a power-law function, $P(k) \propto k^{\gamma}$,
and are very well behaved over the whole
velocity range. We fit a non-weighted power-law function.  To estimate the uncertainty 
for each fit, we iteratively remove one data point and refit. 
The uncertainty on the power-law slope is then calculated as the 
weighted standard deviation of the derived slopes. 
The weighted average slope for these six panels is $\left< \gamma \right > = -3.58\pm0.09$.

\begin{figure*}
\includegraphics{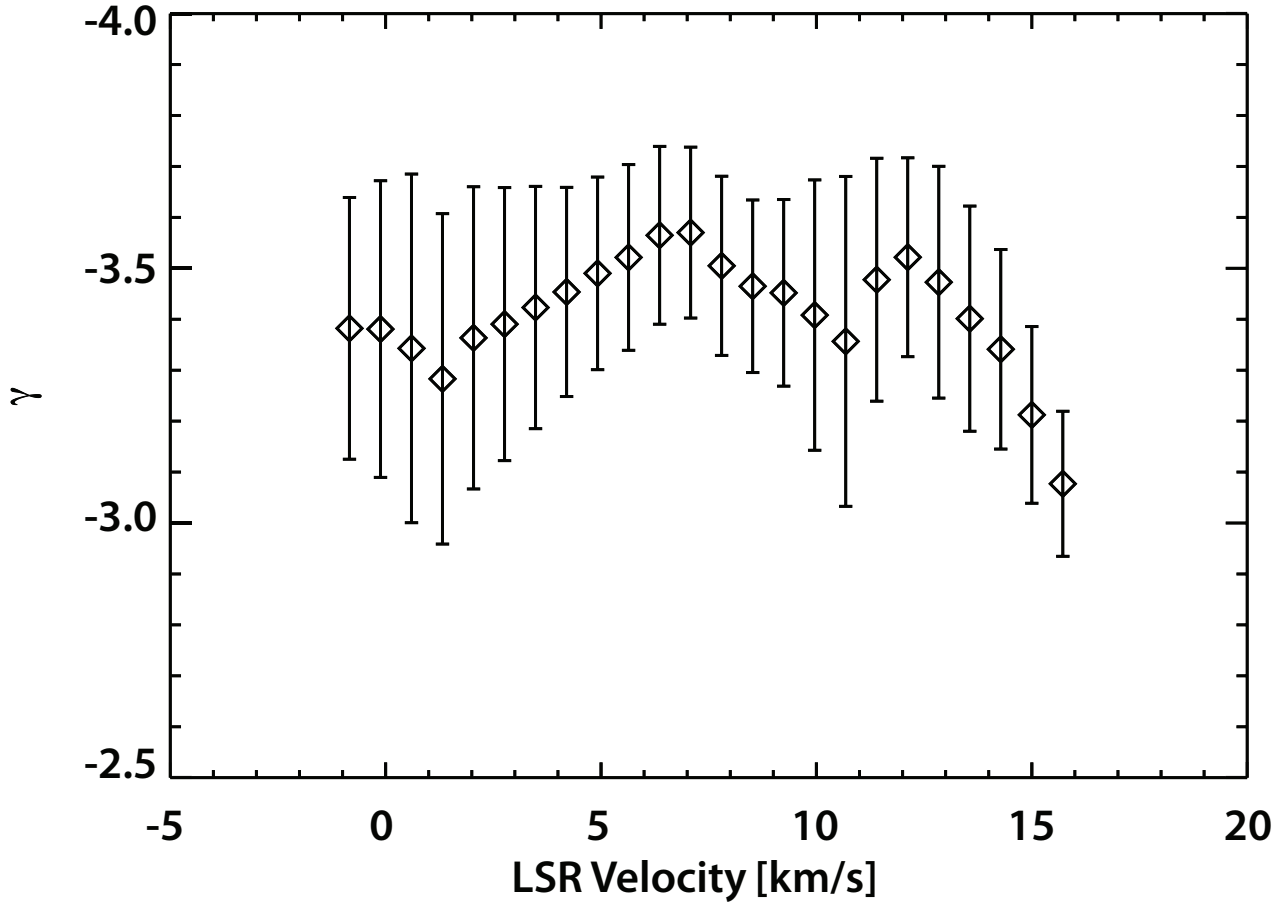}
\caption{The power spectrum slope vs. the LSR velocity of MBM16 for 24 velocity slices each with
$\Delta v=0.72$ \kms. 
Note the relatively constant slope over the velocities corresponding to the bulk of emission.}
\label{fig:fig8}
\end{figure*}

To investigate whether the power-law slope varies with individual velocity channels,
for a given (fixed) velocity channel width, 
we show in Figure~\ref{fig:fig8} the power-law slope as a function of the LSR velocity measured for velocity 
thickness $\Delta v=0.72$ \kms, which correspond to twenty-four individual velocity slices. Within the estimated fitting uncertainties, the power-slope
is essentially constant for a fixed velocity thickness.

As suggested in \citet{lap00}, hereby referred as LP00, intensity fluctuations seen in the H{\small I} images result from a combination of
both velocity and density fluctuations, commonly referred to as $P_v \propto k^{-3-m}$ and 
$P_s \propto k^{n}$.  By integrating a spectral-line data cube over a 
varying range of velocity intervals (or channels), the relative contributions 
of velocity and density fluctuations are expected to vary in a systematic way, allowing 
derivation of power-law indices corresponding to density and velocity fields separately ($n$ and $m$).
Due to this dependence of intensity fluctuations on both velocity and density fluctuations, 
the power-law index is expected to be a function of velocity slice thickness, 
i.e. $\gamma$($\Delta$$v$). At small $\Delta$$v$ the velocity fluctuations 
have a considerable effect on the measured power-law index,  resulting from velocity flows pushing emission 
into adjacent velocity channels \citep{mul04}. However, LP00 showed that these 
velocity fluctuations can be averaged out by integrating over wider velocity
 ranges. At the largest $\Delta$$v$ the velocity fluctuations tend to be averaged out, leaving 
only the contributions from the underlying three-dimensional density structure 
contributing to the measured power-law index. 

\begin{figure*}
\includegraphics{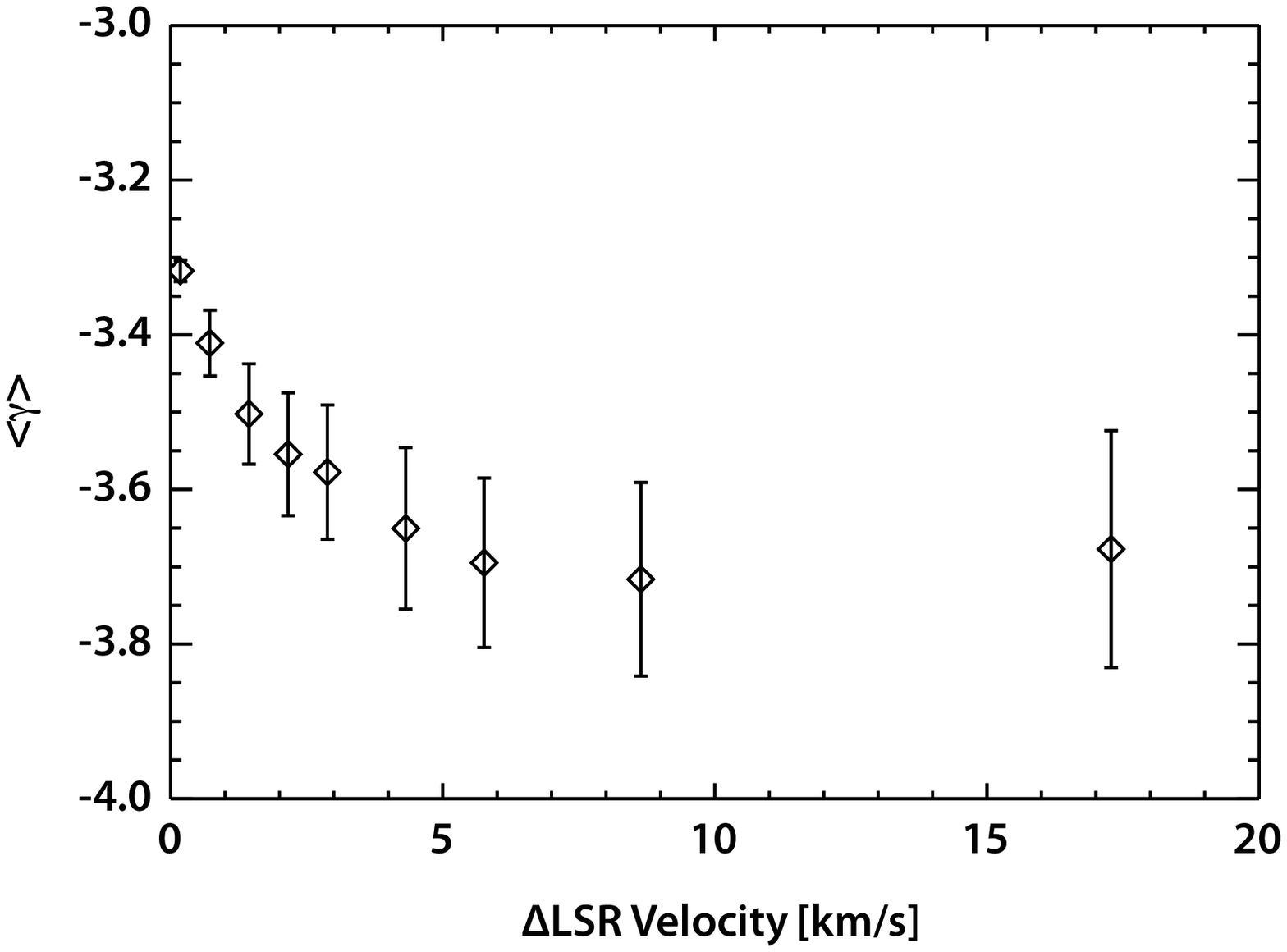}
\caption{Variation of the two-dimentional HI power spectrum slope, $\left<\gamma\right>$, 
with velocity slice thickness $\Delta$$v$.}
\label{fig:fig9}
\end{figure*}

Based on the above expectations, 
we investigate how the power-law slope depends on the thickness of our velocity channels $\Delta$$v$. We start with the full data cube (with individual channels having $\Delta v=0.18$ \kms), 
calculate the power spectrum and its power-law slope for
each individual velocity channel, calculate the weighted-mean slope for each channel, and then gradually average
several velocity channels together to increase  $\Delta v$ all the way to the whole cube being compressed
to a single velocity channel of  $\Delta v\sim18$ \kms. 
We then plot the weighted average with 1$\sigma$ uncertainties as a function of $\Delta$$v$ in 
Figure~\ref{fig:fig9}.

It is clear  that the average power-spectrum slope becomes gradually steeper as velocity slice 
thickness increases, and flattens around $\Delta v\sim5-10$ \kms. 
Based on LP00, this suggests that both 
velocity and density fluctuations contribute 
to the intensity fluctuations in the intensity images, especially at narrow velocity channels. 
This observed trend is in agreement with the LP00 predictions, and
suggests the 3D slope of density fluctuations $n=-3.7\pm0.2$.
The measured SPS slope at the smallest velocity width contains information about
velocity fluctuations. LP00 predict that $\gamma$($\Delta$$v$) should also converge
or flatten at this end of velocity slice thickness.
However, we do not observe this flattening in Figure~\ref{fig:fig9}. This suggests that
even our smallest $\Delta v\sim0.18$ \kms is still not probing the predominantly ``thin'' velocity slices
largely dominated by velocity fluctuations. 
Therefore, based on LP00, if the most shallow slope we measure
corresponds to an intermediate velocity slice thickness case then $\gamma$ = $-3-m/2 \sim -3.3$
and  we get $m\sim0.6$ and the 3D velocity spectrum
$P_v \propto k^{-3.6}$. However, if this slope corresponds to a ``thin'' slice case, then 
$\gamma$ = $-3+m/2 \sim -3.3$ and  we get
$m\sim-0.6$ and $P_v \propto k^{-2.4}$.

\section{Discussion}

The spatial power spectrum of H{\small I} associated with MBM16 shows a power-law behavior 
with a relatively constant power-law slope across the cloud 
velocity range and for a fixed velocity channel thickness. 
This is similar to power spectra derived 
for the Galaxy \citep{cad83,gree93}, for the SMC \citep{sta99,sal01}, 
and the Magellanic Bridge \citep{mul04} in that they are 
well fit by a single power-law function.

The slope of the power spectrum varies with the thickness of velocity 
channels, in agreement with the LP00 predictions. 
The estimated 3D slope of density fluctuations is
steep at $n$ = $-3.7\pm0.2$, essentially identical to the Kolmogorov $-11/3$ 
slope expected for incompressible
turbulence with energy injection on large scales only and dissipation at small scales.
Our rough estimate of the 3D velocity spectrum suggests $P_v \propto k^{\textrm{-2.4 to -3.6}}$
(for the case of Kolmogorov turbulence $P_v \propto k^{-3.7}$). However, as we do not
detect the saturation of the power-spectrum slope at small $\Delta v$ our velocity slope is
not well determined. 
As the power spectra do not show evidence for a turnover on large spatial scales,
this suggests that turbulence is being driven on scales larger than 17 pc.

The most interesting result from this analysis is that
the estimated 3D slope of density fluctuations in MBM16 ($n= 3.7\pm0.2$) is significantly 
steeper than the slope obtained for several regions in the Milky Way plane \citep{cad83,gree93,kha06}, 
the SMC, the Magellanic Bridge, and several other galaxies
(e.g. for the SMC $n=-3.30\pm 0.01$, while the slope of the velocity spectrum is $\sim-3.4$).
The difference could result from the lack of turbulent energy injection at small scales from stellar feedback 
in MBM16, therefore providing evidence that stellar feedback 
provides important contribution to turbulent energy in the ISM.

Previous studies of MBM16's velocity autocorrelation (Magnani, LaRosa, \& Shore 1993; 
LaRosa, Shore, \& Magnani 1999) from CO observations revealed 
evidence for transient coherent structure in the molecular gas 
and suggested that an external shear flow, arising at the 
boundary layer between MBM16 and the surrounding ISM, powers 
the turbulent velocity field in this high-latitude cloud.
The lack of observed turnover in the H{\small I} power spectrum agrees with the scenario that
turbulent energy is being introduced on the spatial scales larger than 17 pc in MBM16. 
Being non self-gravitating and located about $\sim50$ pc above the Milky Way plane,  
MBM16's interstellar environment resembles conditions found in outer galaxy disks
where shear due to Galactic rotation can initiate the magnetorotational instability,
which in turn drives turbulence on large scales (Piontek \& Ostriker 2007).
Alternatively, as suggested and modeled by Keto \& Lattanzio (1989), collisions
between high-latitude clouds can induce gravitational instability and turbulence needed to
also explain broad CO linewidths commonly observed in these clouds.  

The estimated spectral indices agree very well with results from 
another high-latitude field, in Ursa Major,
by Miville-Deschenes et al. (2003) who derived $-3.6\pm0.2$ for the slope of 
the H{\small I} column density fluctuations over the range of scales from 0.5 to 25 pc.
In addition, they measured the same slope for the power spectrum of the H{\small I} velocity 
centroid field. While our velocity fluctuation spectrum is in 
rough agreement with results from Chepurnov et al. (2010) for an adjacent 
high-latitude field, our density slope is significantly steeper.

We  can also compare our derived SPS slope with the prediction 
from isothermal MHD simulations. Figure 10 of \citet{burk10} plots the SPS slope for density fluctuations 
derived from isothermal simulations as a 
function of sonic Mach number $\mathcal{M}_s$ for
super-Alfvenic $(\mathcal{M}_A$ =0.7) and 
sub-Alfvenic ($\mathcal{M}_A$=2.0) simulations. 
The SPS slope becomes more shallow with increasing $\mathcal{M}_s$. 
For any given $\mathcal{M}_A$
the slope is steeper for sub-Alfvenic turbulence.
However, regardless of the super-Alfvenic or sub-Alfvenic case, 
a very steep slope of $\sim-3.7$
is expected only in the case of sub-sonic or transonic medium with $\mathcal{M}_s<2$.
While MBM16 certainly has a mixture of warm and cold H{\small I} phases, 
if the warm medium dominates in terms of its fraction,
then the observed H{\small I} power spectrum suggests a sub-sonic or transonic H{\small I} in MBM16.

In summary, we have used the two-dimensional SPS on high-resolution H{\small I} 
observations to investigate the internal turbulence of the non-starforming, translucent 
molecular cloud, MBM16. 
By fixing the velocity slice thickness we find that the H{\small I} power spectrum is well
represented with a power-law function,
with the slope $\gamma$ staying relatively
constant throughout the full velocity range for the cloud.
By treating $\gamma$ as a function of velocity slice thickness, we find that the average value of
the slopes is influenced by velocity fluctuations at small $\Delta$$v$,
while at large $\Delta$$v$ the influence from velocity has been averaged out to
leave density fluctuations as the dominant mechanism for the observed intensity
fluctuations. The slope of H{\small I} density fluctuations is estimated as $n=-3.7\pm0.2$,
which is steeper than the density spectral index obtained for the Galaxy, the SMC, and the Magellanic Bridge. 
This result for a starless environment suggests that stellar feedback may play a prominent a role in
turbulent energy injection on spatial scales ~0.1-10 pc.
It will be very interesting to see if future studies of starless clouds support our finding.

\acknowledgements
The authors would like to thank Steve Shore for providing the CO observations of MBM16, and
an anonymous refree for providing constructive suggestions.
N.P., S. S., M.-Y.L., M.E.P., C.H., E.J.K.,
J.E.G.P., A.B., J.G., and D.S. acknowledge support from 
NSF grants AST-0707597, 0917810, 0707679,
and 0709347. N.P., M.-Y.L. and S.S. thank the Research Corporation 
for Science Advancement for their support.
J.E.G.P. was supported by HST-HF-51295.01A, provided by NASA
through a Hubble Fellowship grant from STScI, which 
is operated by AURA under NASA contract NAS5-26555. 
We credit the use of the KARMA
visualization software package\citep{karma96}\footnote{See also
\url{http://www.atnf.csiro.au/karma}.}." visualization software (Gooch 1996).

\end{document}